\documentclass[11pt]{article}
  
\usepackage{amsmath,amssymb}

\usepackage{epsfig}  
\usepackage{graphicx}               
\usepackage{url}
\usepackage{hyperref}

\usepackage{pstricks}
\usepackage{color}

\newcommand{\nc}{\newcommand}  

\nc{\beq}{\begin{equation}}  
\nc{\eeq}{\end{equation}}  
\nc{\beqa}{\begin{eqnarray}}  
\nc{\eeqa}{\end{eqnarray}}  
\nc{\bea}{\begin{eqnarray}}  
\nc{\eea}{\end{eqnarray}}  
\nc{\barray}{\begin{eqnarray}}
\nc{\earray}{\end{eqnarray}}
\nc{\barrayn}{\begin{eqnarray*}}
\nc{\earrayn}{\end{eqnarray*}}
\nc{\ra}{\rightarrow}  
\nc{\lsim}{\begin{array}{c}\,\sim\vspace{-21pt}\\< \end{array}}  
\nc{\gsim}{\begin{array}{c}\sim\vspace{-21pt}\\> \end{array}}  
\nc{\Tr}{{\rm Tr}}
\nc{\slsh}{\slash\hspace*{-0.22cm}}

\def\be{\begin{equation}}
\def\ee{\end{equation}}
\def\bea{\begin{eqnarray}}
\def\eea{\end{eqnarray}}
\def\bit{\begin{itemize}}
\def\eit{\end{itemize}}

\nc{\infinity}{\infty}
\nc{\mc}{\mathcal}
\nc{\M}{\mathcal{M}}

\def\to{\rightarrow}

\title{  
\vspace*{-2.3cm}  
\begin{flushright}  
\normalsize{  
SLAC-PUB-15220\\
  }  
\end{flushright}  
\vspace{1.5cm}  
\Large  
\textbf{
Gamma Lines without a Continuum: \\ \vspace{2mm}
Thermal Models for the Fermi-LAT 130 GeV Gamma Line
}\vspace*{1.0cm}  
}

\author{Yang Bai$^{a,b}$ and  Jessie Shelton$^{c}$
\vspace{5mm}
\\
$^{a}$ \normalsize\emph{SLAC National Accelerator Laboratory, 2575 Sand Hill Road, Menlo Park, CA 94025, USA}  \vspace{1mm} \\ 
$^{b}$  \normalsize\emph{Department of Physics, University of Wisconsin, Madison, WI 53706, USA}  \vspace{1mm} \\ 
$^{c}$ \normalsize\emph{Department of Physics, Sloane Laboratory, Yale University, New Haven, CT 06520, USA} 
}

\date{} 

\begin{document}  
\setcounter{page}{0}  
\maketitle  

\vspace*{1cm}  
\begin{abstract} 
  Recent claims of a line in the Fermi-LAT photon spectrum at 130 GeV
  are suggestive of dark matter annihilation in the galactic center
  and other dark matter-dominated regions. If the Fermi feature is
  indeed due to dark matter annihilation, the best-fit line
  cross-section, together with the lack of any corresponding excess in
  continuum photons, poses an interesting puzzle for models of thermal
  dark matter: the line cross-section is too large to be generated
  radiatively from open Standard Model annihilation modes, and too
  small to provide efficient dark matter annihilation in the early
  universe.  We discuss two mechanisms to solve this puzzle and
  illustrate each with a simple reference model in which the dominant
  dark matter annihilation channel is photonic final states. The first
  mechanism we employ is resonant annihilation, which enhances the
  annihilation cross-section during freezeout and allows for a
  sufficiently large present-day annihilation cross section. Second,
  we consider cascade annihilation, with a hierarchy between $p$-wave
  and $s$-wave processes.  Both mechanisms require mass
  near-degeneracies and predict states with masses closely related to
  the dark matter mass; resonant freezeout in addition requires new
  charged particles at the TeV scale.
\end{abstract}  
  
\thispagestyle{empty}  
\newpage  
  
\setcounter{page}{1}  
    
\baselineskip18pt   

\vspace{-3cm}

\section{Introduction}
\label{sec:intro}

The existence of dark matter (DM) is one of the strongest pieces of
evidence for physics beyond the Standard Model.  Searches in cosmic
rays for evidence of DM annihilation or decays are a
cornerstone of the experimental effort to detect DM. Monochromatic
photon lines, though in most models a subdominant signal, provide a
particularly clean astrophysical signal~\cite{Bergstrom:1988fp,
  Bern:1997ng, Bergstrom:2004nr, Bertone:2009cb, Bertone:2010fn}.

Several recent analyses have claimed evidence for a distinct spectral
feature in the Fermi-Large Area Telescope (LAT) \cite{Atwood:2009ez}
photon spectrum at around 130
GeV~\cite{Bringmann:2012vr, Weniger:2012tx, Tempel:2012ey, Su:2012ft}, in
regions near the galactic center. Evidence for this feature has also
been reported in galactic clusters \cite{Hektor:2012kc} and in
non-associated sources \cite{Su:2012zg}, although the latter claim
remains contentious \cite{Hooper:2012qc, Mirabal:2012za, Hektor:2012jc}.
While the statistics are limited, the morphology of the signal may
favor an explanation as annihilating DM
\cite{Cotta:2011pm, Buchmuller:2012rc, Hektor:2012kc}. The presence of
an additional photon line at approximately 111 GeV
\cite{Su:2012ft, Su:2012zg} is highly suggestive, if also statistically
limited, and would lend more credence to a particle physics
explanation \cite{Rajaraman:2012db}.
The Fermi collaboration's own search for photon lines uses slightly
different search regions and methodology and sets an upper limit
marginally in conflict with the claimed
signal~\cite{Ackermann:2012qk}.

It remains to be established whether the excess is instrumental,
astrophysical, or representative of an overly optimistic
characterization of the systematic uncertainties in the galactic
background \cite{Boyarsky:2012ca, Aharonian:2012cs}.  However,
standard thermal WIMPs are not capable of explaining the Fermi signal,
and it is of interest to work out the necessary structure in DM models
which could give rise to the 130 GeV line.

Any dark matter model for the Fermi 130 GeV line must account for
two interesting facts. First, there is no evidence for an excess in
the continuum photon spectrum, which strongly constrains DM
annihilation into the usual SM annihilation channels ($f\bar f$,
$W^+W^-$, $ZZ$)~\cite{Buchmuller:2012rc, Cohen:2012me, Cholis:2012fb,
  Huang:2012yf} or indeed into any charged final states.  Since in
most thermal models DM-photon couplings are generated radiatively from
the DM couplings to charged final states~\cite{Cline:2012nw,
  Das:2012ys}, the typical line cross section is generically related
to the cross-section for annihilation into charged modes $X,X^\dag$ by
\beq 
\label{eq:continuum}
  \langle \sigma v\rangle_{\gamma\gamma} \sim
            \left(\frac{\alpha}{\pi}\right)^2 \langle \sigma  v\rangle_{X X^\dagger}\,.
\eeq
As the fragmentation and decay of the final states $X,X^\dag$ give
rise to a continuum photon spectrum $d\Phi_\gamma (E)/dA \propto
n_{DM}^2\langle \sigma v\rangle_{X X^\dagger} dN_\gamma/dE $, if
annihilation to charged states is open, the expected line flux is
smaller than the continuum flux by several orders of magnitude.
Models for the Fermi 130 GeV line must therefore explain the absence
of annihilation into charged (or hadronic) modes.

This brings us to the second interesting fact.  The best fit
cross-sections for the 130 GeV feature
\cite{Bringmann:2012vr,Weniger:2012tx} are more than an order of
magnitude smaller than the expectation for ($s$-wave) thermal
freezeout.  If annihilation to SM or charged modes is absent or
suppressed as the continuum limits suggest, reconciling this
inefficient annihilation to photons with the WMAP relic density
$\Omega_\chi h^2 = 0.1109\pm 0.0056$~\cite{Larson:2010gs} requires
either: (1) a nonthermal relic abundance~\cite{Acharya:2009zt,
  Tulin:2012uq}; (2) non-photonic annihilation modes which are
suppressed, possibly only post-freezeout, relative to the naive
radiative scaling of Eq.~(\ref{eq:continuum})~\cite{Weiner:2012cb,
  Tulin:2012uq}; or (3) a mechanism to enhance the thermal
annihilation cross-section to photonic final states in the early
universe~\cite{Griest:1990kh}.

In the present work we will study two mechanisms which give enhanced
DM annihilation to photons while also obtaining the correct thermal
relic abundance, and build simple reference models for both.  Our
first example is resonant freezeout, where the presence of a resonance
in the DM sector spectrum enhances the annihilation cross-section into
photons during freezeout. Our second example introduces an
intermediate annihilation mode, so that DM annihilation proceeds
through a cascade decay of a non-photonic intermediate state,
$\bar{\chi}\chi\to \phi \phi' \to 4 \gamma$. In this example, the
relation between the cross-section necessary for thermal freezeout and
the present-day gamma line cross section is explained by the interplay
of $s$-wave and $p$-wave contributions to the annihilation.

In section~\ref{sec:effective}, we perform an effective operator
analysis of DM-photon couplings and demonstrate the need for new
particles at mass scales comparable to the DM mass.  In
section~\ref{sec:resonance} we perform a detailed examination of
resonant freezeout, and consider cascade decays in
section~\ref{sec:cascade}. Section~\ref{sec:conclusion} contains our
conclusions.

\section{Effective Operators for DM Annihilation to Photons}
\label{sec:effective}

To demonstrate the need for multiple new states in thermal models for
the Fermi 130 GeV line, we begin by writing effective operators to
describe the interaction between DM and one or two photons. We assume
an unbroken $\mathbb{Z}_2$ symmetry to explain the stability of the DM
particle, and for simplicity work after electroweak symmetry breaking.
We first consider the case when DM is a (Dirac) fermion.  The leading
operators coupling DM to photons are the electric and magnetic dipole
operators, $\overline{\chi}\sigma_{\mu\nu}\chi F^{\mu\nu}$ and
$\overline{\chi}\sigma_{\mu\nu}\chi \tilde{F}^{\mu\nu}$, appearing at
dimension five~\cite{Bagnasco:1993st, Barger:2010gv}.  However, the
dominant annihilation process mediated by these two operators is
$\chi\chi\to f\bar f$ through a photon in the
$s$-channel~\cite{Weiner:2012cb}.  This problematic annihilation to
$f\bar f$ can be suppressed in the present day if the Dirac fermion is
split into two Majorana $\chi_{1,2}$, with a small mass gap $m_2-m_1$
such that the depleted abundance of the heavier $\chi_2$
post-freezeout shuts off the charged annihilation channel
\cite{Weiner:2012cb, Tulin:2012uq}; in this case, the DM is the
Majorana $\chi_1$ and the EFT containing only $\chi_1$ is indeed
insufficient to describe the freezeout.

At dimension six, there are four operators which couple DM pairs to two
photons, two CP-conserving operators
\beqa
\frac{c_1}{4\,\Lambda^3}\overline{\chi}\chi F_{\mu\nu} F^{\mu\nu} \,+\, \frac{c_5}{4\,\Lambda^3}\overline{\chi}\gamma_5\chi F_{\mu\nu} \tilde{F}^{\mu\nu} \,,
\label{eq:operators}
\eeqa
and two CP-violating operators
\beqa
\frac{\bar{c}_1}{4\,\Lambda^3}\overline{\chi} \chi F_{\mu\nu} \tilde{F}^{\mu\nu} \,+\, 
\frac{\bar{c}_5}{4\,\Lambda^3}\overline{\chi}\gamma_5 \chi F_{\mu\nu} F^{\mu\nu} \,.
\label{eq:operators-violating}
\eeqa
In addition there are operators with a single photon, such as
$\overline{\chi}\chi F_{\mu\nu}Z^{\mu\nu}$,
$\overline{\chi}\gamma_5\chi F_{\mu\nu}\tilde{Z}^{\mu\nu}$, and
$\overline{\chi}\gamma_\mu\chi F^{\mu\nu} \partial_\nu h$. We will
concentrate on the operators with two photons, but we will comment on
other operators when we introduce a concrete UV model.

For scalar DM, the first DM-photon interactions appear at dimension 6,
\beq
\mc{O}_{sc,1} = \phi^\dag \partial_\mu\partial_\nu\phi F^{\mu\nu}, \phantom {spacer} 
\mc{O}_{sc,2} = |\phi|^2 F_{\mu\nu}F^{\mu\nu}, \phantom {spacer} 
\mc{O}_{sc,3} = |\phi|^2 F_{\mu\nu}\widetilde F^{\mu\nu}.
\label{eq:scalars}
\eeq
For simplicity we have taken $\phi$ complex, but this is only
necessary for $\mc{O}_{sc,1}$.  As for the fermionic dipole operators,
$\mc{O}_{sc,1}$ will dominantly mediate annihilation to $f\bar f$,
which can be suppressed by a mass splitting between real and imaginary
parts of $\phi$.

\begin{figure}[h!]
\begin{center}
\hspace*{-0.75cm}
\includegraphics[width=0.48\textwidth]{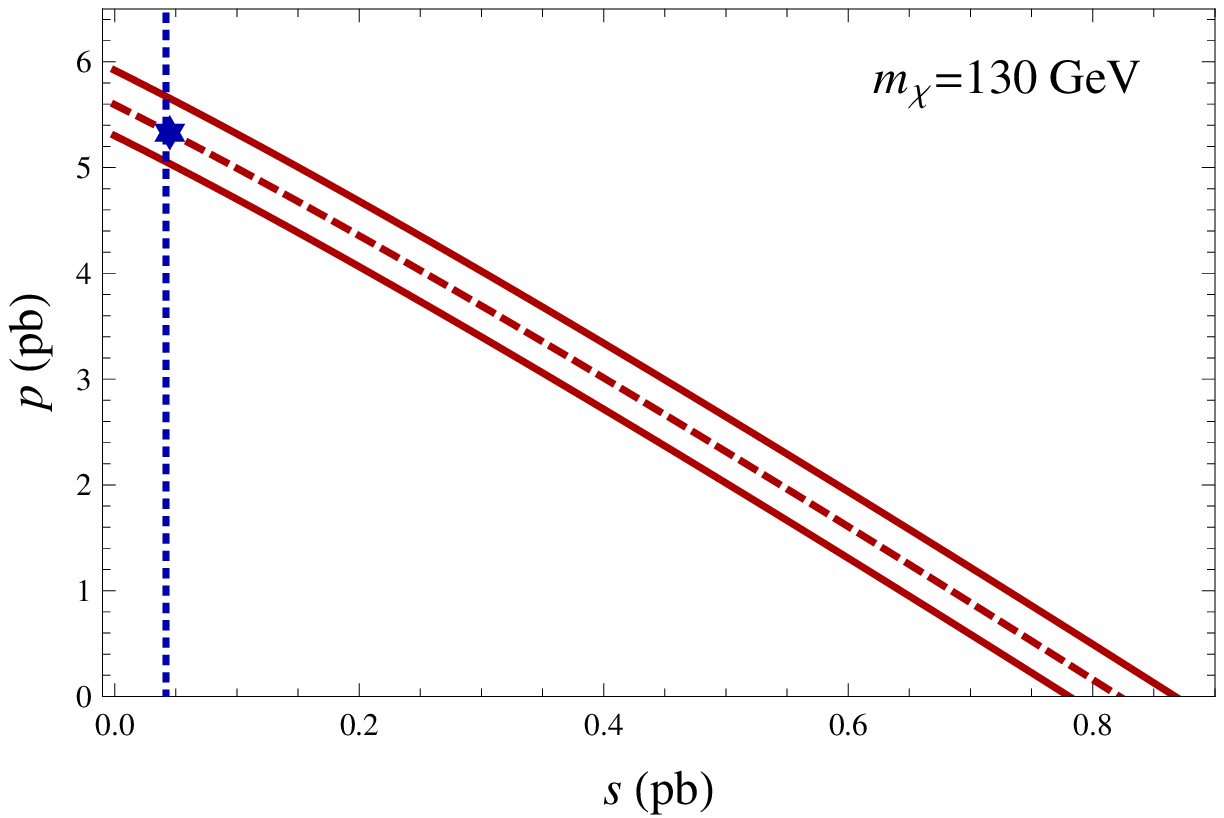} \hspace{3mm}
\includegraphics[width=0.48\textwidth]{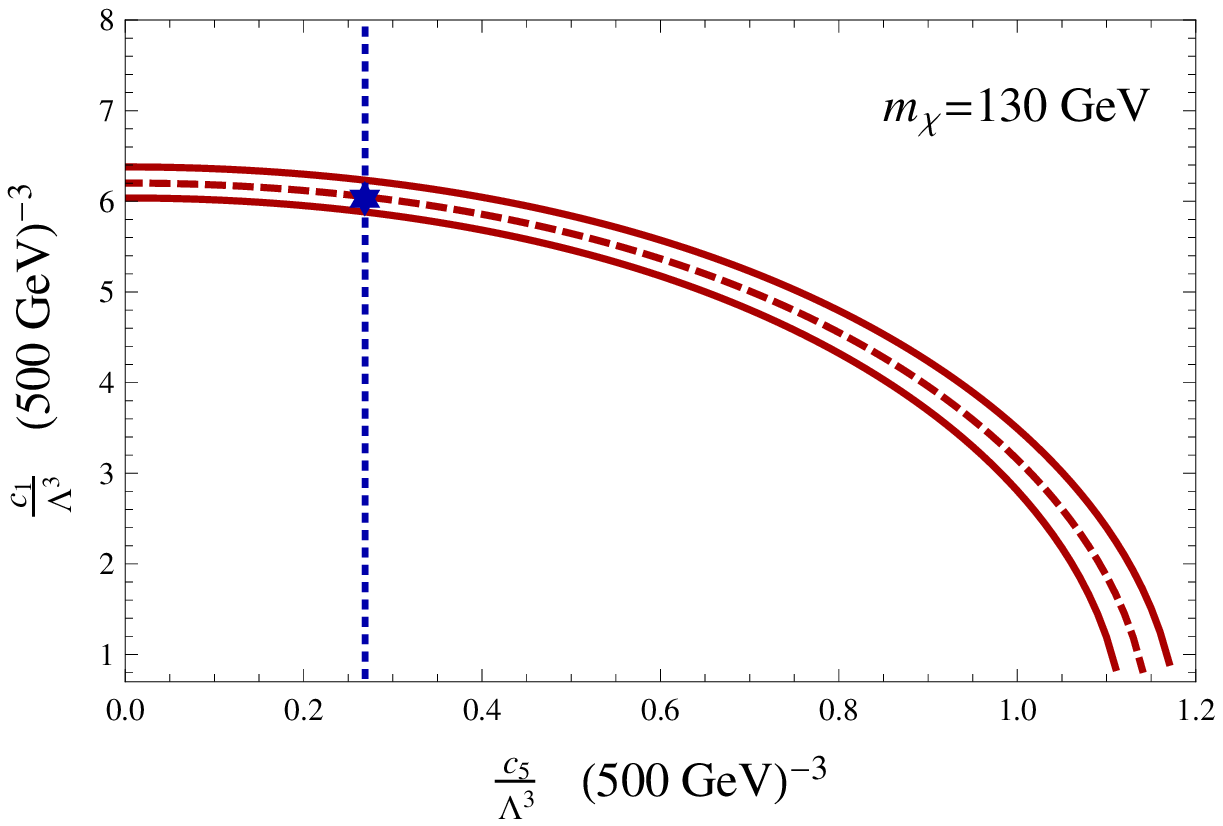} 
\caption{Left panel: the red solid lines are the allowed region of $s$
  and $p$ cross sections in pb to satisfy the dark matter relic
  abundance within one $\sigma$. The blue dotted line denotes
  $s=0.042$~pb$\cdot$c, the present-day gamma line cross section in
  Ref.~\cite{Weniger:2012tx} (Einasto profile). Right panel: same as
  the left, but in terms of operator coefficients $c_1/\Lambda^3$ and $c_5/\Lambda^3$.  The effective number of relativistic degrees of freedom $g_*$ is  taken to be 75.75, $g_{\chi}$ = 4, and $m_{\chi}=130$ GeV.}
\label{fig:relic-parity}
\end{center}
\end{figure}

For fermionic DM, the presence of independent operators which
contribute in different leading partial waves to the DM annihilation
cross-section show that it is easy to accommodate an apparent
suppression in the DM annihilation cross-section.  In an EFT
consisting of the CP-conserving operators in Eq.~(\ref{eq:operators}),
the cross-section is
\beqa
 \sigma v = \frac{m_\chi^4}{16\pi \Lambda^6}\left[
 4\,c_5^2 + (2\,c_5^2 +  c_1^2)\,v^2
\right]  + {\cal O} (v^4) \equiv s + p\,v^2 +  {\cal O} (v^4).
\label{eq:sigmav1}
\eeqa
For a mixed partial wave freezeout process, the dark matter relic
abundance is approximately given by~\cite{Jungman:1995df}
\beqa
\Omega_\chi h^2 \approx \frac{1.07\times 10^9}{\mbox{GeV}\,M_{pl}\,\sqrt{g^*} } \,\frac{x_F}{s + 3( p - s/4)/x_F} \,,
\eeqa
 in terms of the freeze-out temperature
\beqa
x_F = \ln\left[ \frac{5}{4} \sqrt{\frac{45}{8}} \frac{g}{2\pi^3} \frac{M_{pl}\,m_\chi (s + 6\,p/x_F)  }{\sqrt{g^*} \sqrt{x_F} } \right]\,.
\eeqa
In Fig.~\ref{fig:relic-parity} we show in the left panel the region of
$s$ and $p$ giving the correct relic density, and in the right panel
translate that into the region of operator coefficients $c_1/\Lambda^3$ and
$c_5/\Lambda^3$.  From Fig.~\ref{fig:relic-parity}, we can see that to
simultaneously accommodate the thermal relic abundance and the
best-fit cross section for the Fermi-LAT 130 GeV line requires a large
hierarchy between $p$- and $s$-wave scattering, $p/s \approx 130$, or
$c_1/c_5\approx 23$.

Fig.~\ref{fig:relic-parity} also shows that to obtain a reasonable
relic abundance through annihilation to photons only, the cutoffs of
the effective operators are order 500 GeV if $c_1$ and $c_5$ are order of
unity.  Unfortunately the natural magnitude of $c_1$ and $c_5$ is
$\alpha/\pi$, so without additional enhancement factors, the
necessary cutoff $\Lambda$ is much smaller than $m_{\chi}$,
invalidating the EFT.

In the next two sections, we study two models which extend some of the
effective operators of in Eq.~(\ref{eq:operators}) and
Eq.~(\ref{eq:operators-violating}) by introducing a (pseudo-)scalar degree of
freedom $\phi$ which couples to hypercharge gauge bosons through
a loop of charged fermions,
\beqa
{\cal L}=  \bar{f}_i ( i \slsh \partial - g_Y Y B^\mu \gamma_\mu  - m_f ) f_i  - \lambda_\chi^r \bar{\chi}\chi \phi - \lambda_\chi^i i\bar{\chi}\gamma_5\chi \phi
- y_f \phi \bar{f}_i f_i  - \frac{m_\phi^2}{2} \phi^2  \,,
\label{eq:UVmodel-lag}
\eeqa
where $Y$ is the hypercharge of the new charged fermions $f_i$ under
$U(1)_Y$, and $i=1\cdots N_f$ is the flavor index of the $N_f$ fermions.
We integrate out $f_i$ to generate the following effective
operators~\cite{Kniehl:1995tn}
\beqa
\frac{N_f\,y_f\,Y^2\,\alpha_Y}{4\pi\,m_f}\,\frac{2}{3}\,\phi\,B_{\mu\nu}B^{\mu\nu} \supset 
\frac{N_f\,y_f\alpha\,e_f^2}{4\pi\,m_f}\,\frac{2}{3}\,\phi\,F_{\mu\nu}F^{\mu\nu} 
\equiv \frac{\alpha}{4\pi\,\hat m_f}\,\phi\,F_{\mu\nu}F^{\mu\nu} 
\,.
\eeqa
Here we have absorbed the electric charge $e_f$, the Yukawa coupling
$y_f$ and the multiplicity factor $n_f$ into the definition of $\hat
m_f$~\footnote{If $\phi$ is a pseudo-scalar, coupling as $\phi\,i\,\overline{f}_i \gamma_5 f_i$, one needs to
  replace $\frac{2}{3}\times \phi B_{\mu\nu} B^{\mu\nu}$ by $1\times
  \phi B_{\mu\nu} \tilde{B}^{\mu\nu}$.}.  
There are two more operators
$F_{\mu\nu}Z^{\mu\nu}$ and $Z_{\mu\nu}Z^{\mu\nu}$ generated. They
provide a sub-leading contribution to the dark matter annihilation
cross section because of the $\lesssim \theta_W^2 \sim 0.05$
suppression factor.  We neglect these operators in our analysis.  We
also point out that if the $\sim 114$~GeV gamma line becomes robust,
our analysis should be extended to include operators constructed from
$W^a_{\mu\nu}$ to fit the relative fluxes in $\bar \chi\chi\to
\gamma\gamma$ and $\bar{\chi}\chi \rightarrow \gamma Z$.

\section{Resonant Freezeout}
\label{sec:resonance}

One way to enhance the dark matter annihilation cross section at
freezeout relative to the cross-section today is by introducing a
resonance with mass slightly above twice the DM mass,
\beq
m_a^2 = 4 m_\chi^2(1+\delta)\,,
\label{eq:defintion-delta}
\eeq
for $0< \delta \ll 1$ \cite{Griest:1990kh, Ibe:2008ye}.  In this
section we will discuss the parameter space for a resonant freezeout
model where DM annihilation to photon pairs is entirely responsible
for setting the thermal abundance.
 \begin{figure}[t!]
\begin{center}
\hspace*{-0.75cm}
\includegraphics[width=0.45\textwidth]{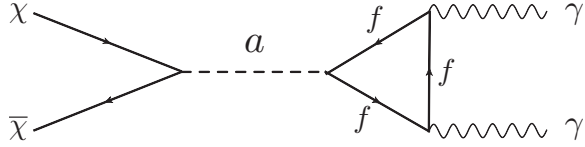}
\caption{Feynman diagram for resonant annihilation.}
\label{fig:fdiagreso}
\end{center}
\end{figure}
To be concrete, we consider a simple UV completion of the EFT in the
previous section, namely Dirac fermionic dark matter $\chi$ coupling
to photons through a pseudo-scalar $a$ with
\beq
{\cal L} \supset - i\,\lambda^i_\chi \overline{\chi}\gamma_5 \chi a  - 
    \frac{1}{4 \Lambda} a \,F_{\mu\nu} \widetilde{F}^{\mu\nu}
 \,.
 \label{eq:resonance-lag-s}
\eeq 
We take the loop-induced $a F\widetilde F$ coupling to be given by
a loop of charged fermions, as shown in Fig.~\ref{fig:fdiagreso},
so the effective coupling is
\beq
\frac{1}{\Lambda} = \frac{\alpha y_f N_f e_f^2}{\pi m_f}\equiv \frac{\alpha}{\pi \hat m_f}\,,
\eeq
where $y_f$ is the Yukawa coupling of $a$ to the heavy fermions $f$
whose mass, number, and charge are given by $m_f$, $N_f$, and $e_f$.

\subsection{$s$-wave}
\label{sec:s-wave}

The operators in Eq.~(\ref{eq:resonance-lag-s}) yield the annihilation cross section 
\begin{eqnarray}
\sigma v &  = & 
    \frac{m_\chi^4}{4\,\pi} \left(\frac{\alpha \lambda_\chi^i}{\pi \,\hat{m}_f} \right)^2 \,
       \frac{1}{\left[ (4m_\chi^2 + m_\chi^2 v^2 ) - m^2_a \right]^2 + \Gamma^2_a m^2_a} \nonumber \\
         & \equiv & \frac{1}{64\,\pi}\left(\frac{\alpha \lambda_\chi^i}{\pi \,\hat{m}_f} \right)^2
                     \frac{1}{(\delta-v^2/4)^2 +\gamma^2(1+2\delta)}  
\end{eqnarray}
where $\Gamma_a$ is the total width of $a$, $\gamma \equiv
\Gamma_a/m_a$, and in the second line we have taken $v,\delta,\gamma\ll
1$ ~\cite{Griest:1990kh}.  We can write the thermally averaged
cross-section as
\beq
\langle\sigma v\rangle \equiv\langle \sigma v\rangle_\infty\, f(x;\delta,\gamma),        
       \label{eq:sigmaavg}
\eeq
where $x \equiv m_\chi /T$,
\beq
\langle \sigma v\rangle_\infty = \frac{1}{64\pi}\left( \frac{\alpha\lambda_\chi^i}{\pi \hat m_f}\right) ^ 2 \frac{1}{\delta^2+\gamma^2(1+2\delta)} 
\eeq
is the cross-section at $x\to\infty$, and
\beq
f(x;\delta,\gamma) =  \frac{x^{3/2}}{\sqrt{4\pi}} \int\, v^2 dv\, e^{-xv^2/4} \frac{\delta^2+\gamma^2(1+2\delta)}{ (\delta+ v ^ 2/4) ^ 2+\gamma^2(1+2\delta)}
\label{eq:s-wave-fx}
\eeq
contains the information about the nontrivial velocity dependence of
the cross-section.  The width of the pseudo-scalar is bounded from
below by its couplings to dark matter (in the relevant parameter
space, the radiative width into photons is negligible). For $\delta\ll
1$,
\beq
\label{eq:gammadef}
\gamma =\frac{ \sqrt{\delta}\lambda_\chi^{i\,2} }{8\pi}(1-\frac{1}{2}\delta)\,.
\eeq

We first establish that our model has a reasonable range of parameter
space which can give a sufficiently large cross-section for
$\chi\bar\chi \to \gamma\gamma $.  Requiring $\langle \sigma
v\rangle_0 = \langle \sigma v\rangle_\infinity f(x_0)$,
\beq
\left(\frac{64 \pi^3 \hat m_f^2}{\alpha^2\lambda_\chi^{i\,2}}\right) \langle \sigma v\rangle_0 =   \left(\frac{\hat m_f /\lambda_\chi^i}{\mathrm{300\;GeV}}\right)^2  \left(\frac{1}{0.06} \right)^2 = 
      \frac{1}{\delta^2} f(x_0;\delta) \approx \frac{1}{\delta^2}\,,
      \label{eq:delta-mf}
\eeq
where we have used the Einasto value for $\langle \sigma v\rangle_0$
and dropped terms of order $\gamma^2$.  This determines $\hat m_f$ as
a function of $\delta$ and $\lambda_\chi^i$, showing that obtaining
the present-day cross-section requires $\delta\lesssim 0.1$.

\begin{figure}
\begin{center}
\hspace*{-0.75cm}
\includegraphics[width=0.48\textwidth]{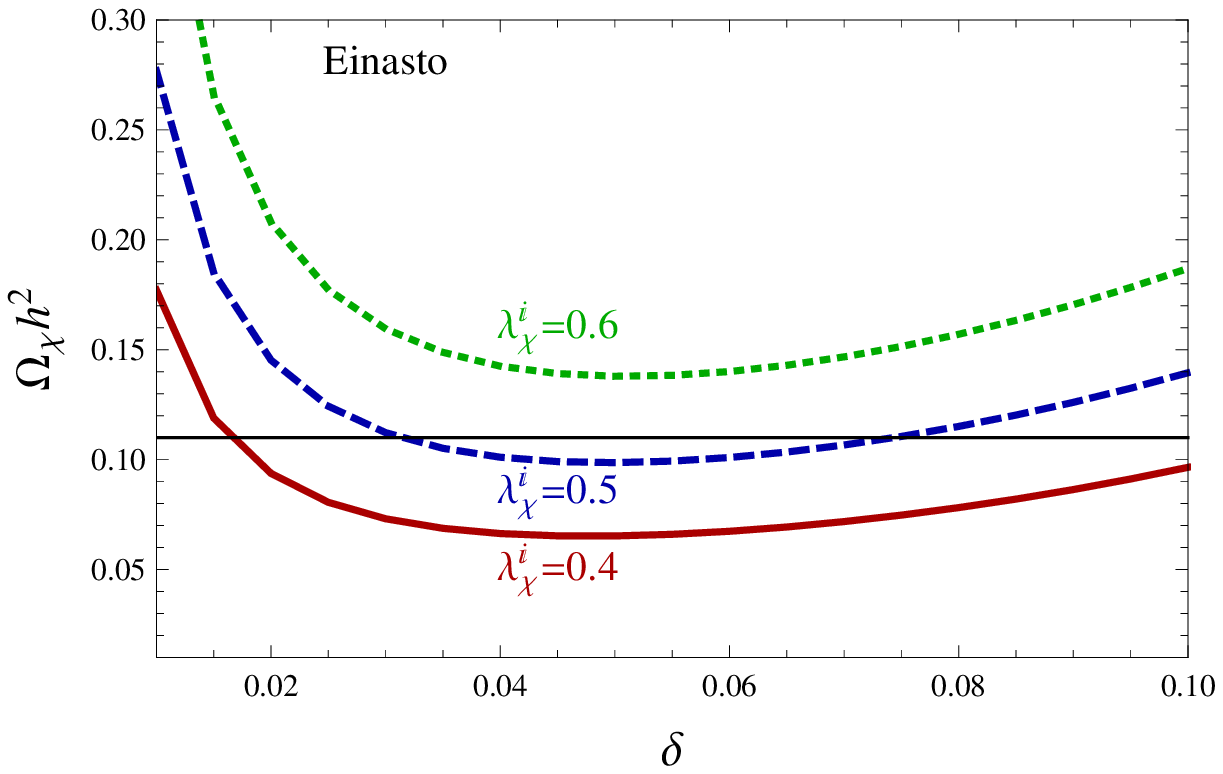} \hspace{3mm}
\includegraphics[width=0.48\textwidth]{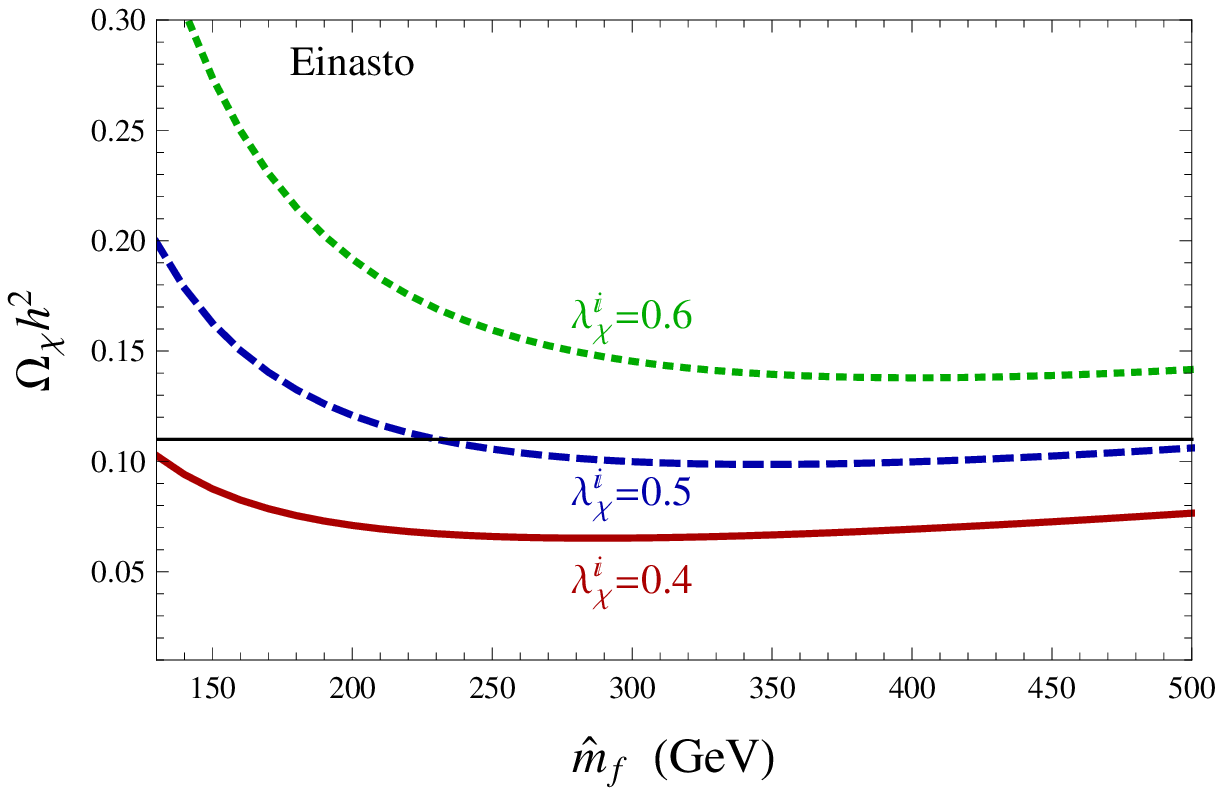}
\caption{The dependence of the relic abundance $\Omega_{\chi}h^2$ on
  $\delta$ for different values of $\lambda_\chi^i$ (left) and $\hat
  m_f$ (right), with the present-day cross-section fixed to the
  (Einasto) best-fit value for the Fermi-LAT gamma line excess.}
\label{fig:deltacurves}
\end{center}
\end{figure}

After using the present-day best fit cross-section to fix $\hat m_f$
in terms of $\lambda_\chi^i$, $\delta$, the thermal freezeout is
controlled by $\delta$ and the minimum allowable width, as determined
by Eq.~(\ref{eq:gammadef}).  Figure~\ref{fig:deltacurves} illustrates
 the dependence of the final relic abundance on $\delta$ and
$\lambda_\chi^i$. 
Since the resonance sees significant overlap with the bulk of the
velocity distribution during freezeout, the strength of the resonant
enhancement to the cross-section is enhanced when $\gamma$ is smaller,
increasing the value of the cross-section at the pole.  Therefore if
$\gamma$ is too small, e.g.~$\lambda_\chi^i \le 0.4$ in Fig.~\ref
{fig:deltacurves}, the annihilation is too efficient and the yield is
too small to account for the present-day dark matter abundance in the
absence of other decay modes for $a$.  If, on the
other hand, $\gamma $ is too large, e.g.~$\lambda_\chi^i \ge 0.6$ in Fig.~\ref
{fig:deltacurves}, annihilation is inefficient and could even over-close the
universe.  Annihilation becomes less efficient both gradually at large
$\delta$ and rapidly at very small $\delta$, as the pole passes
outside the main peak of the Maxwell-Boltzmann distribution during
freezeout.

 \begin{figure}[h!]
\begin{center}
\hspace*{-0.75cm}
\includegraphics[width=0.48\textwidth]{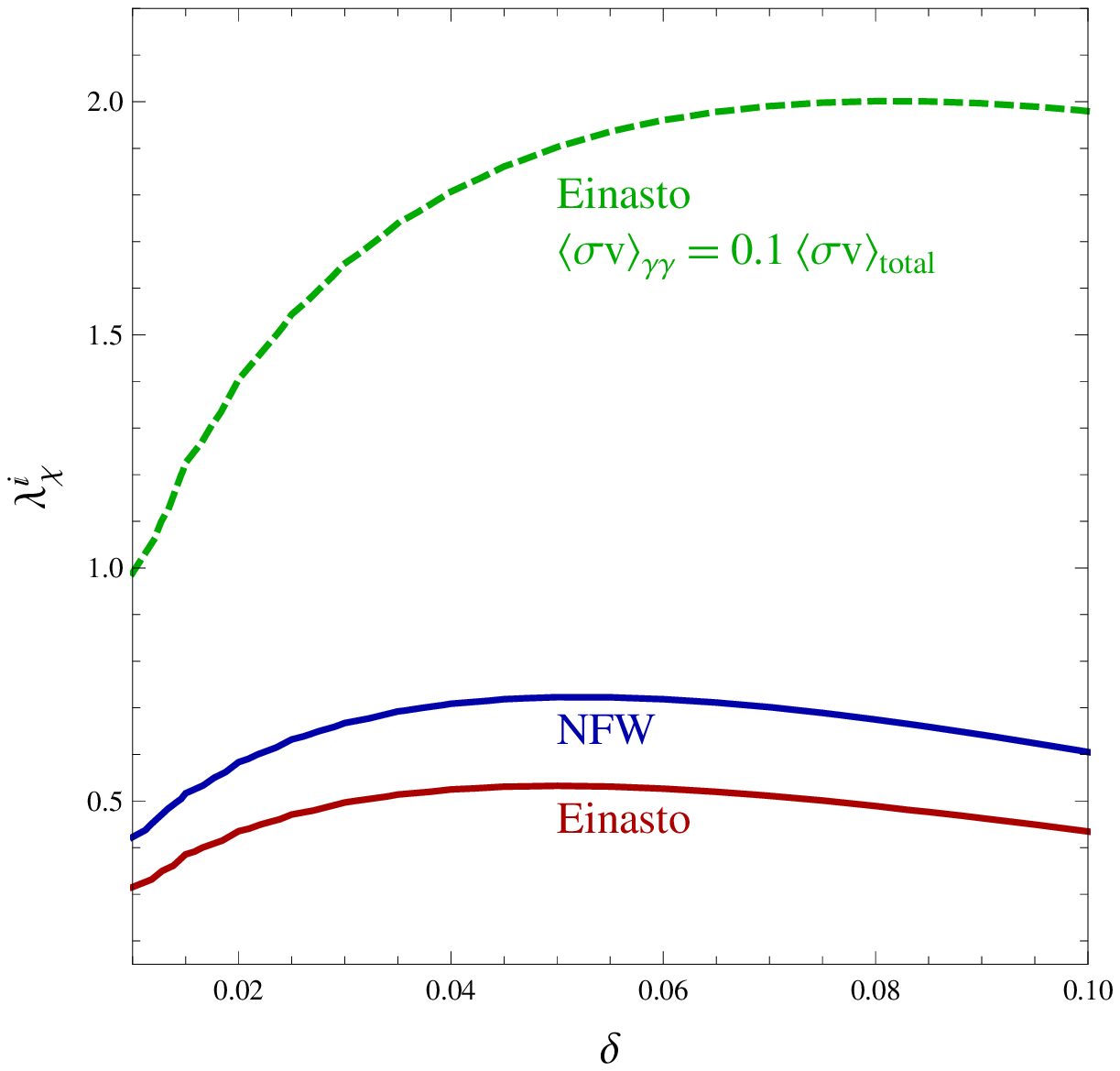} 
\hspace{3mm}
\includegraphics[width=0.48\textwidth]{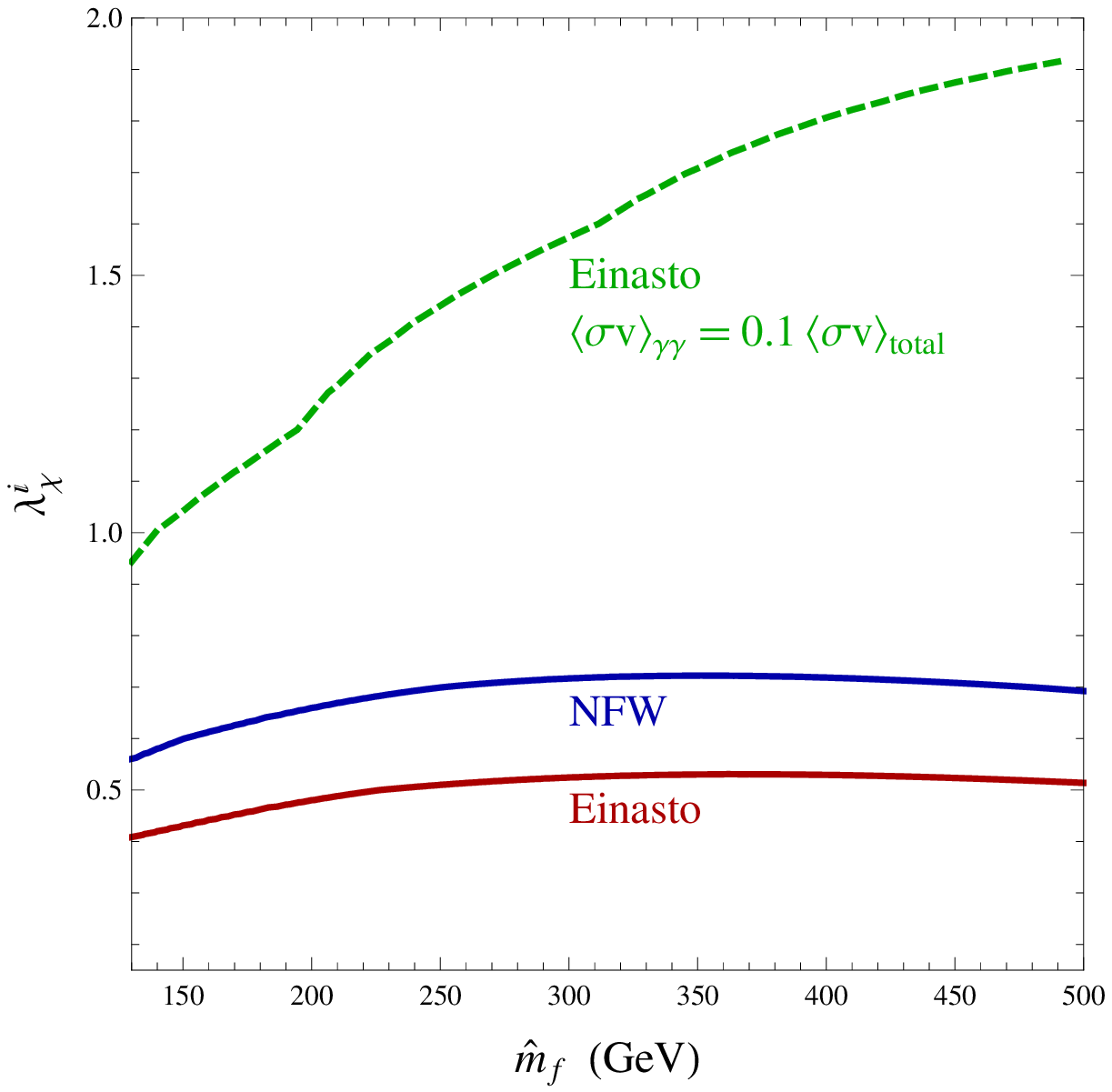} 
\caption{Contours in the $\delta$-$\lambda_\chi^i$ plane (left) and  $\hat
  m_f$- $\lambda_\chi^i$ plane (right)  yielding $\Omega_{\chi} h^2 = 0.111$
  after fixing present-day cross-sections to best-fit Fermi-LAT gamma
  line values.   The dashed line is the contour for the present-day
  Einasto gamma line cross section with the assumption that $\mbox{Br}(a\rightarrow \gamma\gamma)=10\%$.} 
\label{fig:deltalambda}
\end{center}
\end{figure}

In Fig.~\ref{fig:deltalambda} we show the contours in the
$\delta$-$\lambda_\chi^i$ plane yielding $\Omega_{\chi} h^2=0.11$.  To
illustrate the astrophysical uncertainties, we quote results for both
the Einasto and NFW gamma line best-fit cross sections. Comparing the
resulting contours, we can see that the astrophysical uncertainty
introduces an order 50\% uncertainty on the couplings of the
pseudo-scalar to dark matter. In the green dashed line of
Fig.~\ref{fig:deltalambda}, we show the contour for the case where $a$
has an additional decay mode $XX^\dag$ with $\mbox{Br}(a\to XX^\dag) = 10\,
\mbox{Br}(a\to\gamma\gamma)$.

As can be seen from Fig.~\ref{fig:deltalambda}, there is no real upper
bound on the (scaled) fermion mass $\hat m_f$. However, from
Eq.~(\ref{eq:delta-mf}), $\hat m_f$ is inversely proportional to
$\delta$, to obtain the present-day annihilation cross section. If
one requires $\delta > 0.01$ from considerations of fine-tuning, the
scale for the charged fermion masses should be bounded from above by
around 500 GeV, or in other words, within collider-accessible energies.

\subsection{$s+p$-wave}

We now consider the case where more than one partial wave is important
for freezeout.  For concreteness we extend the single  pseudo-scalar model
 of the previous subsection to
include a CP-violating coupling $\lambda_\chi^r a
\bar\chi\chi $ to the dark matter,
\beq
\mc{L}\supset 
 - i \lambda_\chi^i \overline{\chi}\gamma_5 \chi a -  \lambda_\chi^r a \bar\chi\chi - 
     \frac{\alpha}{4\pi \hat m_f} a \,F_{\mu\nu} \widetilde{F}^{\mu\nu} .
\eeq

Pure $p$-wave freezeout would require extremely tuned values of the
resonance mass, $\delta \sim 10^{-5}$, to obtain the present-day
observed cross-section.  In the more interesting case of mixed $s$ and
$p$-wave resonant freezeout, the present-day observed cross-section,
and hence the necessary values of $\delta$, are set by the $s$-wave
contribution, but $p$-wave scattering can dominate the annihilation at
freezeout.  When $p$-wave scattering dominates freezeout, the
additional $v^2$ dependence of the cross-section shifts the velocity
integral to higher values, reducing the overlap with the pole;
narrower resonances are therefore required to achieve cross-sections
sufficiently efficient to yield the observed relic abundance.
Therefore when $p$-wave scattering dominates during freezeout,
additional contributions to the $a$ width are more constrained than in
the $s$-wave case.  Since we will be interested in cases with a
hierarchy between $p$-wave and $s$-wave modes, it is important to
include the contribution to $\gamma$ from $\lambda_{\chi}^r$,
$\gamma_r =  \delta^{3/2}\lambda_\chi^{r\,2}(1-\frac{1}{2}\delta) /(8\pi)$.

 \begin{figure}[h!]
\begin{center}
\hspace*{-0.75cm}
\includegraphics[width=0.48\textwidth]{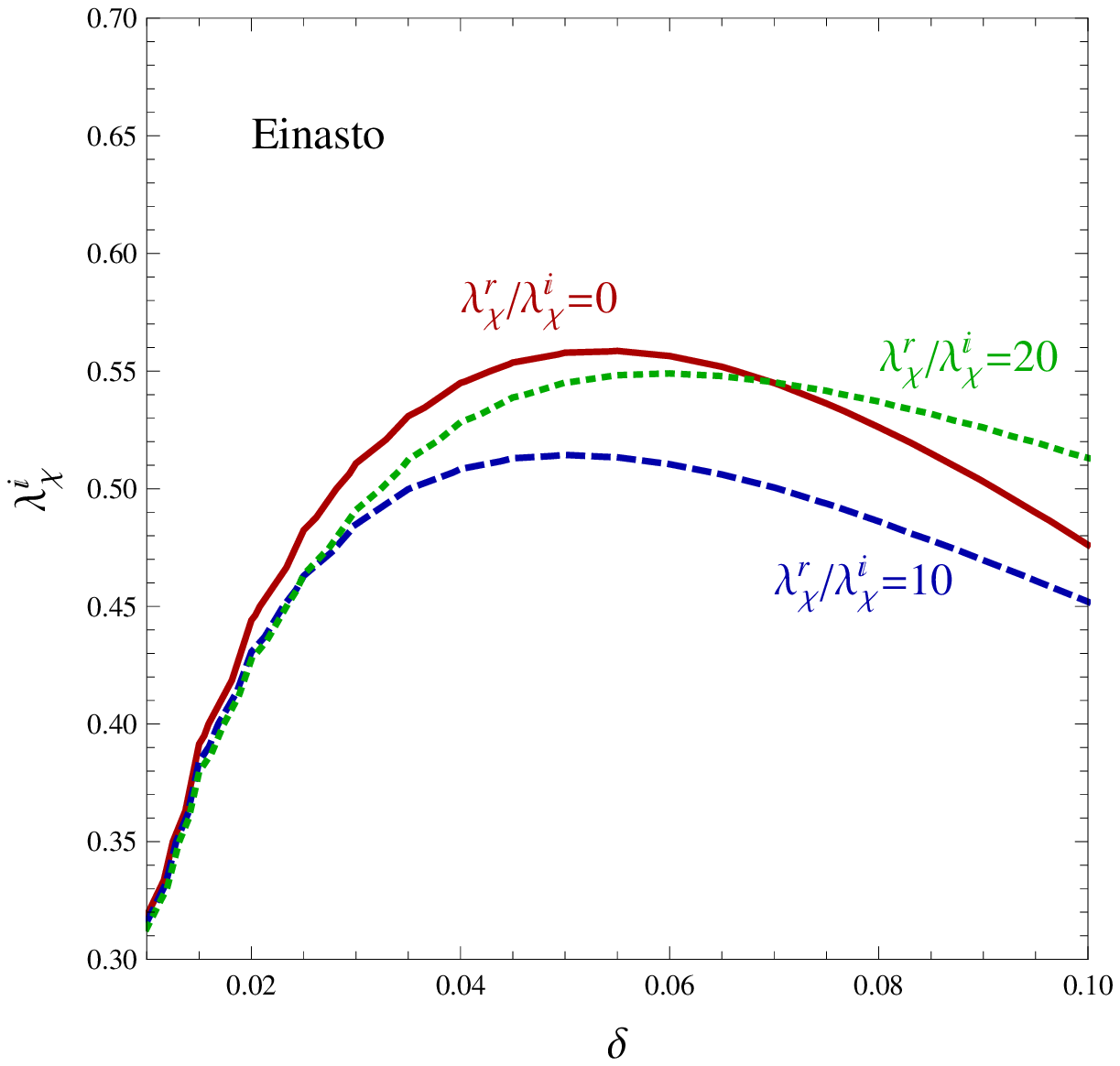}
\hspace{3mm}
\includegraphics[width=0.48\textwidth]{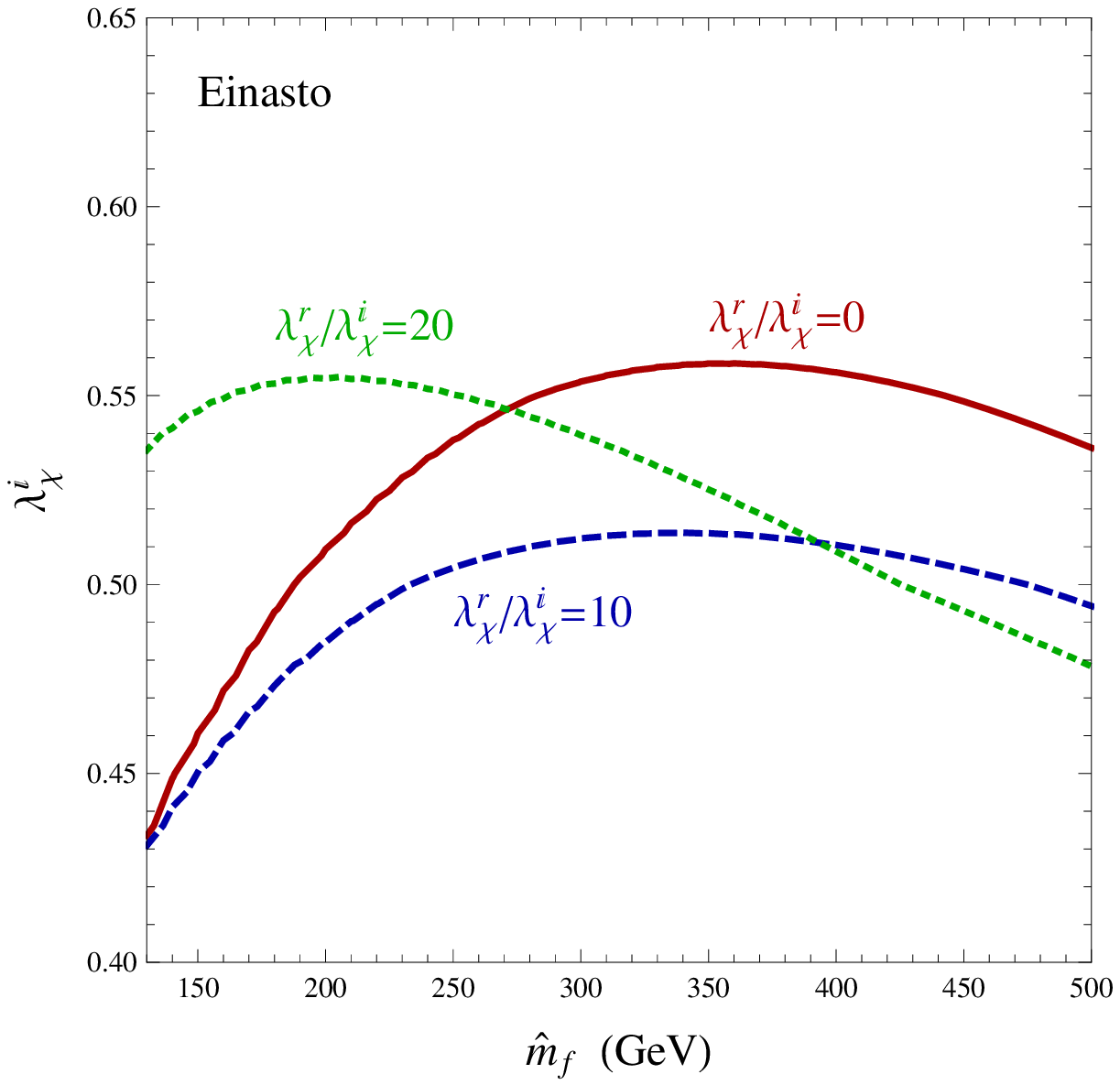}
\caption{Left panel: contours in the $\delta$-$\lambda_\chi^i$ plane
  yielding $\Omega_{\chi} h^2 = 0.11$, for different ratios of
  $\lambda_\chi^r/\lambda_\chi^i$.  Results are shown for the the
  Einasto best fit cross-section for the gamma line excess. Right
  panel: the same as the left one but in the $\hat
  m_f$-$\lambda_\chi^i$ plane.}
\label{fig:deltalambdaSP}
\end{center}
\end{figure}
As for the pure $s$-wave case, we show the contours of $\Omega_{\chi}
h^2= 0.11 $ in the $\delta$-$\lambda^i_\chi$ and $\hat
m_f$-$\lambda_\chi^i$ planes in Fig.~\ref{fig:deltalambdaSP}.  We plot
three different ratios of $\lambda_\chi^r/\lambda_\chi^i$. For
$\lambda_\chi^i \gtrsim \lambda_\chi^r \sqrt{\delta}$, the
contribution to the resonance width from $\lambda_\chi^r$ is
negligible. The additional $p$-wave contribution to the freeze-out
cross section prefers a smaller value of $\lambda_\chi^i$, which
explains why the pure $s$-wave contour (red, solid) in the left panel
lies at larger values of $\lambda_\chi^i$ than the blue dashed
contour, which denotes a case where $p$-wave scattering is important
for annihilation but not for the resonant width.  For $\lambda_\chi^i
\lesssim \lambda_\chi^r \sqrt{\delta}$, the contribution to the
resonance width from $\lambda_\chi^r$ is non-negligible, reducing the
enhancement from the resonance. As a result, larger values of
$\lambda_\chi^i$ are required to obtain sufficiently efficient
annihilation, as can be seen from the green, dotted contour in
the left panel.

As can been from the right panel of Fig.~\ref{fig:deltalambdaSP}, the
photon vertex scale $\hat m_f$ is again below around 500~GeV for $\delta
\gtrsim 0.01$, implying collider-accessible charged particles.

\section{Cascade Annihilation}
\label{sec:cascade}

Another way to reconcile the large photon line
cross-section with the lack of continuum photons while obtaining the
proper relic density is to have the dark matter first annihilate into
(neutral) intermediate states, which then decay into photons.  The
advantage of extending the DM annihilation process with a cascade
decay is that the relic abundance is now only controlled by the
couplings of dark matter to the intermediate particles. The small
radiative couplings of photons to the intermediate state are only
relevant for the lifetime of that state, and are irrelevant for the
annihilation cross section. The gamma ray signals from cascading DM
annihilations have been discussed in~\cite{Mardon:2009rc, Bai:2009ka,
  Fortin:2009rq} and applied to the Fermi-LAT gamma line excess
in~\cite{Ibarra:2012dw, Tempel:2012ey}. Here, we write down a simple
explicit model and discuss the parameter space for realizing a thermal
relic abundance together with an explanation for the Fermi-LAT gamma
line excess.

\begin{figure}[t!]
\begin{center}
\hspace*{-0.75cm}
\includegraphics[width=0.45\textwidth]{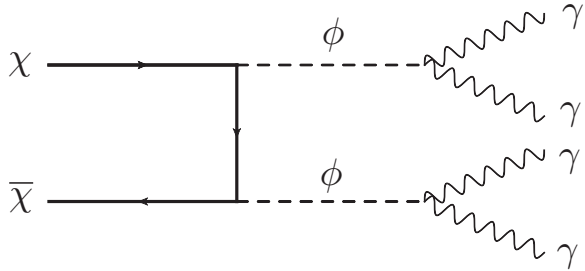}
\caption{The Feynman diagram to describe DM cascade annihilation into four photons.}
\label{fig:feynman-4a}
\end{center}
\end{figure}

As before, we consider Dirac dark matter $\chi$, together with a real
scalar field which we denote $\phi$, and study the following set of
interactions
\beqa
{\cal L} \supset - \lambda_\chi^r \overline{\chi} \chi \phi - i \lambda_\chi^i \overline{\chi}\gamma_5 \chi \phi  - \frac{\alpha}{4\pi\,\hat{m}_f}\phi\,F_{\mu\nu} F^{\mu\nu}
\label{eq:cascade-lag}
 \,.
\eeqa
Here, as before, the scale $\hat m_f\equiv 2 N_f y_f e_f^2/3 m_f$
arises from some heavy charged fermions that are integrated out to
generate the $\phi F F$ operator; we require only that $m_f$ is
sufficiently large to forbid DM from annihilating into charged
fermions.  We are interested in the parameter space where $m_\phi <
m_\chi$, so the dominant annihilation channel for DM is
$\overline{\chi}\chi \rightarrow \phi \phi \rightarrow 4\gamma$, as
shown in Fig.~\ref{fig:feynman-4a}.

The energies of the two photons from each $\phi$ decay are, in the dark
matter rest frame,
\beqa
E_{\gamma_{1,2}} = \frac{m_\chi}{2} \left( 1 \pm \sqrt{1- \frac{m_\phi^2}{m_\chi^2} } \cos{\theta} \right) \,,
\eeqa
where $\theta$ is the angle between the photon direction in the $\phi$
rest frame and the $\phi$ direction of motion.  Because $\phi$ is a
scalar field, the distribution is isotropic in $\theta$, and the
photon spectrum is evenly distributed between the kinematic endpoints:
\beqa
\frac{d N_\gamma}{N_\gamma d E_\gamma} &=& 
    \frac{1}{\sqrt{m_\chi^2 - m_\phi^2} } 
  \Theta\left[ E -  \frac{m_\chi}{2} \left(1 - \sqrt{1- \frac{m_\phi^2}{m_\chi^2}} \right) \right]
  \Theta\left[  \frac{m_\chi}{2} \left(1 + \sqrt{1- \frac{m_\phi^2}{m_\chi^2}}\right) - E  \right]
         \,,  \nonumber \\
& \xrightarrow{\epsilon \ll 1} & \frac{1}{ m_\chi \sqrt{2\epsilon} } 
  \Theta\left[ E -  \frac{m_\chi}{2} \left(1 - \sqrt{2\epsilon} \right) \right]  
  \Theta\left[ \frac{m_\chi}{2} \left(1 + \sqrt{2\epsilon} \right) - E  \right] \,,
\label{eq:mediator-spectra}
\eeqa
which, as $\epsilon\to 0$, limits to a delta function centered at
$m_\chi/2$. Here $\Theta(x)$ is the usual Heavyside function.

 \begin{figure}[h!]
\begin{center}
\hspace*{-0.75cm}
\includegraphics[width=0.55\textwidth]{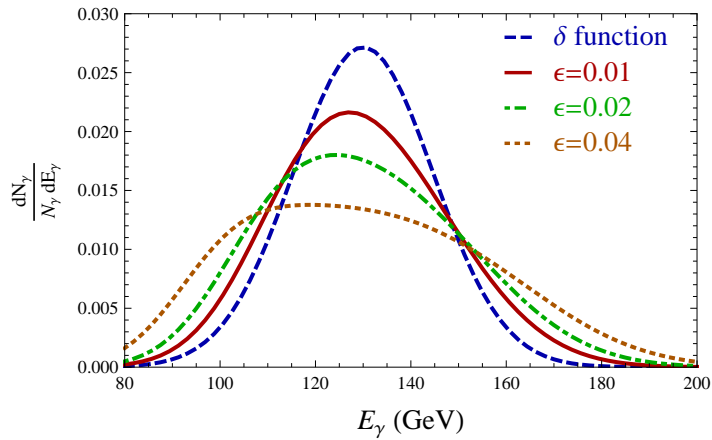}
\caption{The energy-smeared rectangular photon spectra of
  Eq.~(\ref{eq:mediator-spectra}) for different values of the mass
  splitting $\epsilon$. The dark matter mass is chosen to be 260 GeV
  for the red solid, green dotdashed, orange dotted lines. As a
  comparison, we also show a smeared delta function at 130 GeV in the
  blue dashed line.}
\label{fig:rectspectra}
\end{center}
\end{figure}

Since the gamma line spectrum can provide a good fit to Fermi LAT
data, we also anticipate a good fit for this model for sufficiently
small $\epsilon$.  To estimate upper bounds for $\epsilon$, we
consider the average Fermi-LAT energy resolution for gamma ray
energies above 50 GeV, $\sigma(E)/E \approx 0.10 + 0.0001\,E/$GeV
~\cite{FermiResolution}.  We use this energy resolution to smear the
spectra in Eq.~(\ref{eq:mediator-spectra}) for different values of
$\epsilon$ and compare them with a smeared delta function spectrum
centered at 130 GeV. The results are shown in
Fig.~\ref{fig:rectspectra}, where we have shown three different values
of $\epsilon = 0.01, 0.02, 0.04$ in the red solid, green dot-dashed,
and orange dotted lines. The delta function spectrum is shown in the
blue dashed line. As one can see, once $\mc{O}(\sqrt{\epsilon}
/2)$ is smaller than the energy resolution (order 10\%) the
distinctions between the smeared delta function and the smeared cascade
spectra are minimal.

To work out the parameter space for this model we translate the
best-fit line spectra of \cite{Weniger:2012tx} to the cascade model.
Since the photon flux from dark matter annihilation is inversely
proportional to the square of the DM mass, and there are now four
photons in the final state, the required annihilation cross section
for this case should be approximately twice that found for a photon
line. Thus we require $\langle \sigma v \rangle_0 = 0.084$~pb$\cdot$c
for the Einasto profile and $0.152$~pb$\cdot$c for the NFW profile.
 \begin{figure}[h!]
\begin{center}
\hspace*{-0.75cm}
\includegraphics[width=0.50\textwidth]{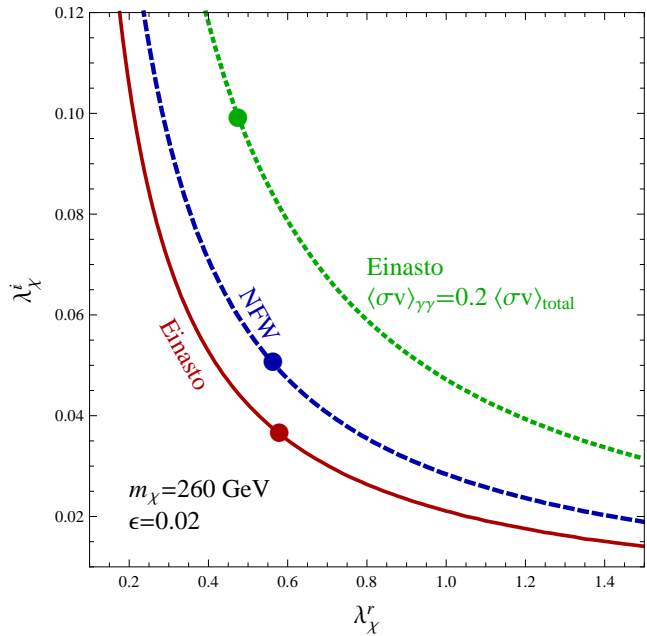}
\caption{The contours of couplings $\lambda_\chi^i$ and $\lambda_\chi
  ^r$ which yield the best-fit cross-sections for the gamma line
  excess. The solid circles indicate the points which yield a thermal
  relic abundance of $\Omega_{\chi}h^2=0.11 $. The green dotted line is the
  case when $\mbox{Br}(\phi\to \gamma\gamma)=0.2$ of the total.}
\label{fig:mediator-parameter}
\end{center}
\end{figure}

Our explicit model considers fermionic DM and, for economy, a single real
scalar $\phi$.  Thus parity dictates that the annihilation is $p$-wave
suppressed unless both $\lambda_\chi^r$ and the CP-violating
$\lambda_\chi^i$ are nonzero.  In terms of the mass splitting
$\epsilon $, the cross-section for $\overline \chi\chi\to \phi\phi$ is
\beqa
\sigma v = \frac{\lambda_\chi^{r\,2} \lambda_\chi^{i\,2} \,\sqrt{\epsilon} }{ \sqrt{2} \pi m_\chi^2 } + v^2\left[  \frac{\lambda_\chi^{r\,2} \lambda_\chi^{i\,2} }{ 16\sqrt{2} \pi m_\chi^2 \, \sqrt{\epsilon}}
+  \frac{ \lambda_\chi^{r\,2}  ( 16 \lambda_\chi^{r\,2} -87 \lambda_\chi^{i\,2} ) \sqrt{\epsilon} }{ 64\sqrt{2} \pi m_\chi^2 \, }
\right] \, + \mc{O}(v^4)
\eeqa
where we have kept the leading terms in the limit $\epsilon \ll
1$.

In Fig.~\ref{fig:mediator-parameter}, we show the contours in the
$\lambda_\chi^i$-$\lambda_\chi^r$ plane which give the required dark
matter annihilation cross section for explaining the Fermi-LAT gamma
line excess, for $m_\chi = 260$~GeV and $\epsilon=0.02$.  The points
which give the relic abundance $\Omega_{\chi} h^2=0.11$ are indicated by
heavy circles.  Hierarchies of order $\sim 10$ between CP-preserving
and CP-violating couplings are required, suggesting small (but not tiny)
CP violation in a DM sector.  We also show the case where
the branching fraction of $\phi$ into two photons is 20\%.

An equally well-motivated alternative to the CP-violating reference
model considered here would be to allow the Dirac DM $\chi$ to
annihilate to a nearly degenerate pair of scalars with opposite parity,
$\phi$ and $a$, with subsequent decays to photons.  In this case, the
$s$-wave annihilation $\overline{\chi}\chi\to \phi\,a$ is proportional
to $(\lambda_\chi^\phi \lambda_\chi^a)^2$, while the $p$-wave
annihilation is proportional to $\sim (\lambda_\chi^\phi)^4 +
(\lambda_\chi^a)^4$.  If the branching fractions of $\phi$, $a$ into 
photons are order 1, accommodating the dark matter thermal relic
abundance and the present-day annihilation cross section for the
Fermi-LAT gamma line still requires a factor of $\sim 10$ hierarchy
between $\lambda_\chi^\phi$ and $\lambda_\chi^a$.

\section{Discussion and Conclusions}
\label{sec:conclusion}

We have explored the possibilities for thermal models for the
Fermi-LAT 130 GeV gamma line excess where the dominant DM annihilation
channel is into photons.  We consider two mechanisms, (1) models where a resonance
in the DM spectrum enhances the annihilation rate during freezeout and
to a lesser extent at the present day, and (2) models where DM
annihilates to a new intermediate state which subsequently decays to
photon pairs.  Here the interplay of the $s$-wave and $p$-wave
annihilation is responsible for reconciling the necessary
cross-section at freezeout with the observed cross-section today.
For the resonance model, charged fermions at the TeV scale are
predicted and are accessible at the Large Hadron Collider, which could be interesting if the excess in the Higgs diphoton decay channel persists~\cite{AtlasHiggs,CMSHiggs}.

Both of the classes of models considered in this paper require
coincidences in the mass spectrum. For resonant freezeout, the
resonance mass $m_a$ must be within a percent of twice the dark matter
mass, a striking coincidence.  For cascade decays, the intermediate
state(s) $\phi$ must be within again about a percent of the dark
matter mass in order to have a sharp enough spectral feature to fit
the data well.  These near-degeneracies are suggestive of a dark
sector with one scale $\Lambda_m$ setting the overall mass scale, and
another $\Lambda_G \ll \Lambda_m$ determining splittings.  Composite dark sectors
are an appealing avenue to flesh out the models we have considered.
Near-degeneracies in the dark sector spectrum, as necessary for the cascade
annihilation model, are readily accommodated in composite sectors.
Resonances above threshold pose a somewhat more complicated picture.
As we know from heavy quarkonia in the SM, a spectrum with resonances
slightly above threshold could certainly be obtained; the complication
is that such resonances would necessarily be accompanied by a
resonance {\em below} threshold, giving the thermal relic abundance a
detailed dependence on the parameters of the model.  We leave detailed
model building to future work.


\subsection*{Acknowledgments} 
We would like to thank Tim Tait, Haibo Yu and especially Simona Murgia
for useful discussions and comments. SLAC is operated by Stanford
University for the US Department of Energy under contract
DE-AC02-76SF00515. JS was supported by the DOE grant DE-FG02-92ER40704
and by the LHC Theory Initiative through the grant NSF-PHY-0969510.
We also thank the Aspen Center for Physics, under NSF Grant
No. 1066293, where this work was completed.

\bibliographystyle{JHEP}
\providecommand{\href}[2]{#2}\begingroup\raggedright\endgroup

 \end{document}